\documentclass[12pt]{article}
\usepackage{amsfonts}
\usepackage{epsfig}
\usepackage{url}
\begin{document}
\title{Exploration-exploitation trade-off features a saltatory search behaviour}

\author{Dimitri Volchenkov${}^1$\footnote{Corresponding author. E-mail address:   volchenk@physik.uni-bielefeld.de}, Jonathan Helbach${}^2$\footnote{Presently with IDIADA Fahrzeugtechnik GmbH,
 Manfred-Hochstatter-Stra{\ss}e 2, D-85055 Ingolstadt-Etting,  Germany}, \\ Marko Tscherepanow${}^2$\footnote{Presently with Beckhoff Automation GmbH, H\"{u}lshorstweg 20, 33415 Verl, Germany}, Sina K\"{u}hnel${}^3$}
 
\maketitle

\begin{flushleft}
${ }^1${\it   Faculty of Physics,  Bielefeld University,  Universitaetsstr. 25, 33615 Bielefeld, Germany} 

${ }^2${\it   Technical Faculty,  Bielefeld University} \\

${ }^3${\it Physiological Psychology, Bielefeld University}
\end{flushleft}

\begin{abstract}
Searching experiments conducted in different virtual environments over a gender balanced group of people revealed a gender irrelevant scale-free spread of searching activity on large spatiotemporal scales. We have suggested and solved analytically a simple statistical model of the coherent-noise type describing the exploration-exploitation trade-off in humans ("should I stay or should I go"). The model exhibits a variety of saltatory behaviours, ranging from Levy flights occurring under uncertainty to Brownian walks performed by a treasure hunter confident  of the eventual success. 
\end{abstract}

\begin{flushleft}
\end{flushleft}

\renewcommand{\baselinestretch}{1.2}
\normalsize

\section{Introduction \label{sec:Intro}}
\noindent

L\'{e}vy foraging hypothesis 
\cite{Bartumeus:2007}-\cite{Viswanathan:2011}
  suggests that
a saltatory search 
composed of consecutive displacement
 lengths $l$ drawn from a power law distribution,
\begin{equation}
\label{Levy}
P(l)\,\,\sim \,\, l^{-\mu},
\end{equation}
with the scaling exponent $\mu$ 
approaching the 
theoretically optimal value $\mu=2$
would maximize forager's chance 
to locate 
sparsely and randomly distributed prey 
\cite{Viswanathan:1996}-\cite{Buldyrev:2001}
 and therefore presents an evolutionary beneficial alternative 
strategy to spatially intensive search.
Foraging movement patterns that 
fit closely to walks (\ref{Levy}) 
known as L\'{e}vy flights  \cite{Levy:1996}
were identified in many living spices 
ranging from microorganisms to humans   
 \cite{Viswanathan:1996}-\cite{Buldyrev:2001}, \cite{Shlesinger:1986}-\cite{Sims:2012}, 
although the reported scaling exponents
 varied substantially
  for different 
animals, in different environmental contexts.
The rate of flight lengths decay
in the population data
 is typically described 
by a power law with an exponential cut-off revealing 
that an exponential decay rate dominates
for extremely long travels 
\cite{Sims:2008, Edwards:2007}.
It was further suggested \cite{Turchin:1996,Petrovskii:2011}
that  while the distribution  
of displacements for the
population aggregate appears to show a fat tail,
the individual's bout distributions do not. 
Furthermore, when
movement lengths within tracks fit a Brownian walk
 (being exponentially distributed), but differ in the parameter of this exponential distribution, 
 the power law with exponential cut-off in the population data would result from a superposition of these exponential distributions
 \cite{Petrovskii:2011,Matthaus:2011}. The detailed 
statistically robust analysis of foraging trajectories of albatrosses
 \cite{Humphries:2012} provided strong support for individual 
character of search patterns: 
some trajectories were well approximated by 
truncated L{\'e}vy flights (\ref{Levy}), 
others fitted Brownian movement patterns, while 
a significant portion of trajectories were
not fitted by either distribution, as having both 
L{\'e}vy and Brownian features.

Further progress in understanding of
search behaviour calls 
for a model of 
a feasible biological mechanism 
that allow animals and humans
i.) reproducing a variety of 
statistically different movement patterns,
without demanding and tedious computations,
and ii.) spontaneous switching between them in different 
environments. In our paper, 
we propose such a model and solve it analytically (Sec.~\ref{Sec:SOC}).

The VE  
provides a simplified way to see and experience 
the real world, 
supporting the sense of spatial presence via virtual locomotion,
rendering a clear sense of  navigation, and
allowing for interactions with objects through a user interface
\cite{Wann:1996}.
Not surprisingly, VE gained widespread use in recent years as a tool for studying human behavior, 
maintaining the capacity to
create unique experimental scenarios 
under tightly controlled stimulus conditions.
A major problem for users of VE is
maintaining knowledge of their location and orientation
while they move through  space, essentially 
when the whole path cannot be viewed at once
 but occluded by objects in the environment \cite{Darken:1993}.
Most of  human spatial abilities (such as 
navigating a large-scale space
and identifying a place)
have evolved in natural environments over a very long time, using 
properties present in nature as cues for spatial orientation and wayfinding
\cite{Werner:2003}.
However, many of natural body-based self-motion cues 
are absent in VE, causing systematic spatial
orientation problems in subjects, and 
therefore calling for  
the instant invention of new adaptive strategies
 to move through VE, 
under reduced multisensory conditions.
The analysis of adaptive strategies instantaneously evolving 
in VE  accentuates the key biological mechanisms 
of searching behaviour more vividly than 
the analysis of empirical data 
recorded in situ.
Furthermore,
 in VE we can study 
 the mobility patterns of humans 
with extremely high resolution,
 up to the scales of millimetres and milliseconds, 
by far obtaining the most accurate data
 virtually comparable to
neither
 the scales of 
a few meters/ tenth of seconds, 
 kilometres/hours
 (in GPS tracking data, \cite{Rhee:2011}),
or the scale of 
a few thousand kilometres / weeks
( in bank note travel patterns, \cite{Brockmann:2006})
discussed in the previous studies.

In order to understand the adaptive movement strategy  and 
to clarify the role of environmental structure 
in searching and browsing, 
we conducted 
a treasure hunting experiment (Sec.~\ref{subsec:Experimental_design})
with a gender balanced group of 
 participants (Sec.~\ref{subsec:Participants}),
 in the different office VE (Sec.~\ref{subsec:Environements}).
People participated in our study 
had to decide 
how to proceed amid uncertainty 
by solving the
{\it exploration-exploitation dilemma} ("should I stay or should I go") \cite{Cohen:2007}.
On the one hand, there was an option to
continue searching  in the nearest neighbourhood ({\it exploitation}), in 
 the hope to get rewarded beyond the next door.
Alternatively, subjects could {\it explore} the parts of  environment 
she never been to. 
The actual trajectory of search 
resulted from a permanent balance 
between exploitation and exploration 
confronted at all levels of behaviour
across all time-scales. 
We show that balancing exploitation
 and exploration can be responsible  
for the statistically variable searching
 behaviour, ranging from L\'{e}vy flights to Brownian walks. 
It is worth a mention that 
the setting of our experiments
was inconsistent with 
the assumptions of  Gittins that
  presented an optimal strategy for trading
off exploration and exploitation \cite{Gittins:1989}.  
The probability of delivering a reward 
was not fixed in our case, subjects did not discount  
the value of each reward  
exponentially as a function of when it was acquired, and eventually
the time of our experiment was essentially limited 
(in contrast to an infinite time horizon in the Gittins' approach \cite{Gittins:1989}). Subjects in our study acted amid uncertainty,
so that any precursive calculation of
 an optimal strategy was impossible for them.

Gender is often reported as a decisive factor in spatial cognition research 
\cite{Lawton:1994,Devlin:1995}. 
A review of gender differences in spatial
ability in real world situations can be found in \cite{Voyer:1995}.
In the present paper,
we do not discuss the gender mobility differences
 observed in our experiments,  
leaving a detailed report on that for a forthcoming publication.
Based on the results of the statistical data analysis, 
we discuss 
on the role of scanning and reorientations in
a compensation of information deficiency
while moving through VE (Sec.~\ref{subsec:reorientations}), 
and on the experimentally observed super-diffusive
spread of searching activity on large spatiotemporal scales
 (Sec.~\ref{subsec:Levy}).

In Sec.~\ref{Sec:SOC},
we have formulated  
a mathematical  model of decision making 
when no precise information on a possibility of rewards is available.
The model can be 
solved analytically 
for some important cases (see Sec.~\ref{subsec:solution}) 
and helps to generate biologically relevant hypothesis
about the fundamental process of  decision making.
We conclude in the last section.

\section{\label{sec:Methods} Methods}
\noindent

\subsection{\label{subsec:Experimental_design} Experimental design and procedure}
\noindent

In our treasure searching experiments, every participant
was asked to browse an office VE searching for 
collectable objects. For 
each time frame, the position and heading orientation
of the participant were tracked and subsequently analysed
by considering the collections of displacements and turns 
as a series of random events whose spatial and temporal distributions are assumed to possess certain 
statistical regularities.

 To motivate 
subjects for searching thoroughly,
 each found object were rewarded 
with an extra 50 cents coin,
 in addition to the basis  
  remuneration  
for participation in the study.
Treasure hunters neither 
 visited a real prototype of the VE model,
 nor foresaw it's floor plan before participating 
in the experiment. 
The objects of search
were big enough, contrast coloured,
 clearly visible  
toys: teddy bears and locomotives. 
At the beginning of each trial, 
a number of toys 
(10 toys, for the smaller environment, and 15 toys, for the bigger one; 
see the Sec.~\ref{subsec:Environements} for details) 
were allocated 
in randomly chosen offices,  
 beyond the closed doors, one toy per room. 
Objects could be found immediately,
 as soon as subject opens 
the door and enters the room. 
In order to focus subjects on the tasks,
 no communication between
experimenter and subject was performed during the experiment.

Before entering the main exploration areas, 
every participant was trained in a virtual tutorial room,
in order to get used to stereoscopic imaging of computer-simulated 
environment (two-slightly different images accounting for the
interpupillary distance paired with stereo glasses providing a 3D display of the environment),
to  get a good command of 
a {\it Nintendo} Wii remote controller,
 and to judge their perceived motions via button presses.
Although the time of search was not limited,
we have restricted the 
total number of doors
subjects could open
during the experiment
(by 10 doors, in the smaller environment, and by 15 doors, in the bigger one), 
in order to prevent a sequential search  
at each office and to stimulate an exploration activity in subjects.
The experiment ended when the participant opened 10 (15) doors.

Two AVI video fragments 
showing the records of actual searching experiment 
from a first-person perspective can be found under the URLs \cite{Movies}.

\subsection{\label{subsec:Environements} Virtual environments}
\noindent

The virtual models of two  office environments
existing in
the University of Bielefeld were rendered
with the {\it Autodesk} $\circledR$ {\it 3ds Max} $\circledR$ {\it Design 2010} software and then projected 
for any user viewpoint
onto a wall-wide laboratory screen (4{\it m}$\times$2{\it m}) with 
the use of the {\it Barco Galaxy NH-12}
active stereoscopic 3D stereo projector.
The sense of spatial presence
was 
 in subjects
supported
in subjects by
natural colour reproduction,
extended gray levels, and high brightness of the projector.
 The control of user
viewpoint motion through the VE 
  was implemented 
via the {\it Bluetooth} connected
{\it  Wiimote}, the primary controller for {\it Nintendo's} Wii console
featured with motion sensing capability, which allows 
the user to  manipulate items on screen via gesture recognition and pointing through the use of accelerometer and optical sensor technology.

\begin{figure}[ht!]
 \noindent
\centering
\begin{tabular}{lll}
A). & \epsfig{file=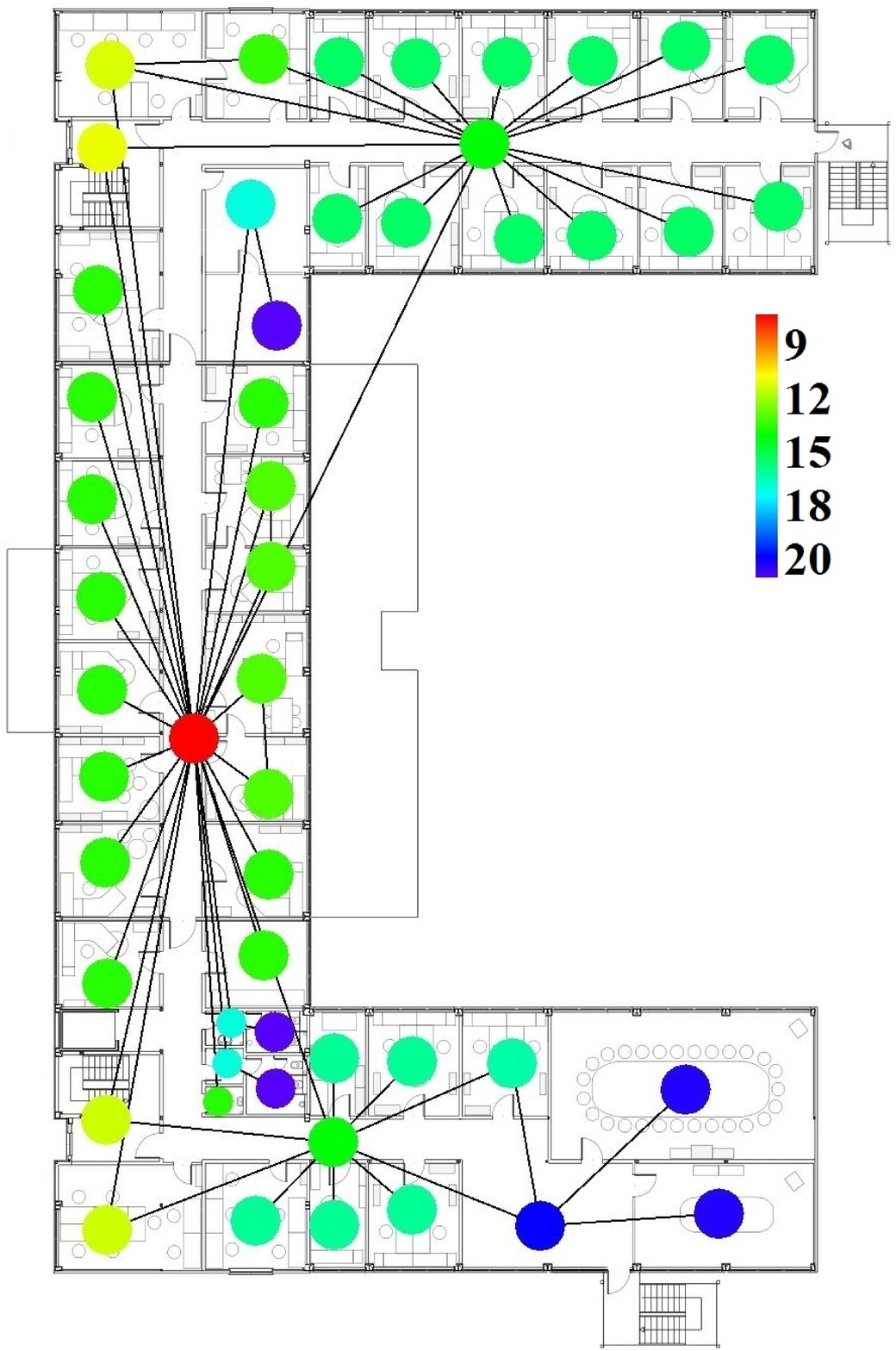, width=4.6cm, height =6cm}
& \epsfig{file=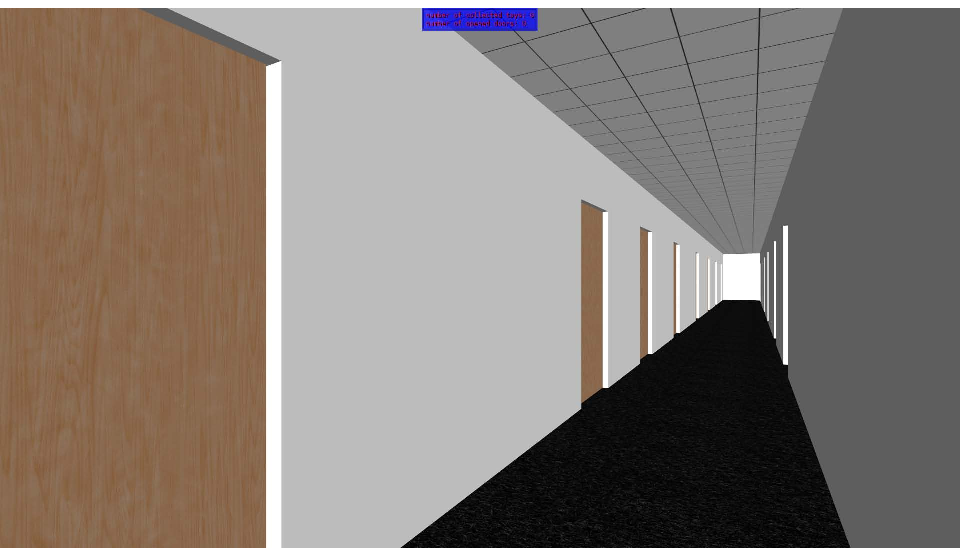, width=6cm, height =4cm} \\
B). &  \epsfig{file=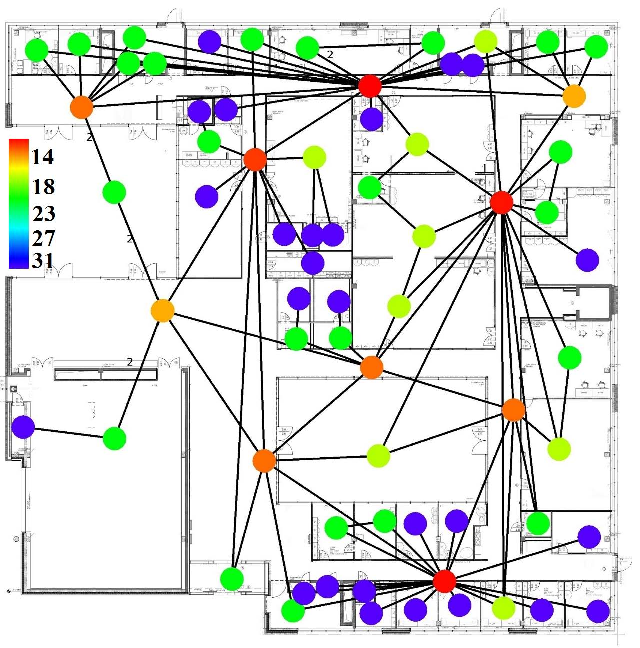, width=6cm, height =6cm} 
 &  \epsfig{file=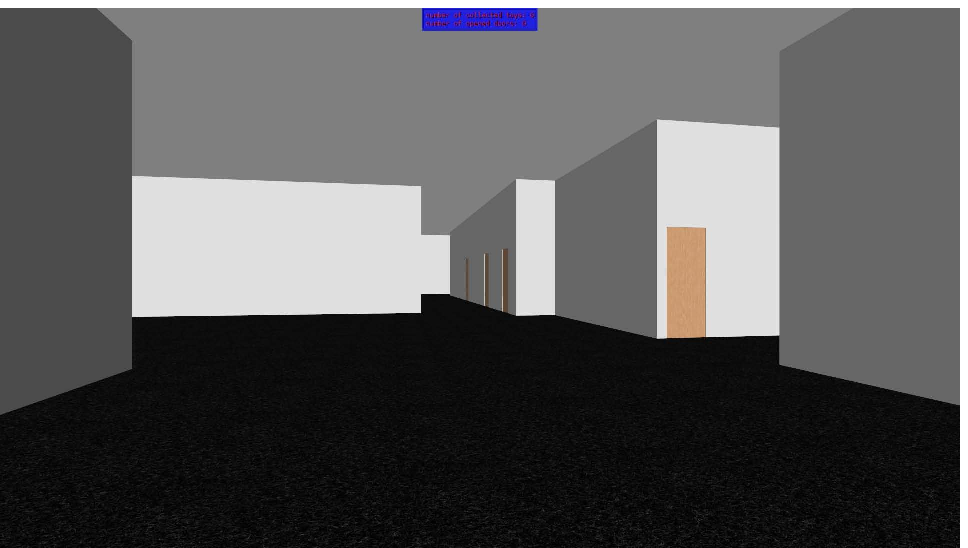, width=6cm, height =4cm}
\end{tabular}
\caption{ (Colour online) Two models of the office VE were used in the treasure search experiments. The individual spaces of movement are identified with the nodes in 
the spatial graphs. The centrality/ isolation of a node with respect to the entire structure of a spatial graph is characterized by the 
first-passage time of random walks indicated by colour.  
A.)  The VE  model A consists of 48 interconnected spaces of movement, 
with a single central place (the central corridor denoted by the red node)
  that locates 9 steps apart from any randomly chosen node
in the spatial graph.
B.) The VE  model B consists of 68 interconnected spaces of movement and
  contains a network of well-connected central places (halls and connecting corridors) characterized  
by the first passage times ranging from 10 to 14 steps. \label{Fig_01}}
\end{figure}

The VE model A (Fig.~\ref{Fig_01}.A)
exactly reproduced 
the second floor of a temporary  
building  (2012)
belonging to the Cognitive Interaction Technology - Centre of Excellence 
(CITEC, Bielefeld University),
 and the VE model B (Fig.~\ref{Fig_01}.B) 
imitated
the ground floor of the future Interactive Intelligent Systems Institute
(Bielefeld University) 
presently under construction.
The both environments consist of the standard
adjacent offices, meeting rooms, hallways and corridors
providing space where people can move, meet, and discuss. 
Emergency exits and elevators that exist in the 
actual  buildings were not taken into account 
in our experiments.
The VE model A consists of 48 interconnected  individual 
spaces of movement
(represented by nodes in the spatial graph shown in Fig.~\ref{Fig_01}.A), 
and the VE model B includes 68 interconnected individual spaces 
of movement
(the nodes of the 
spatial graph shown in Fig.~\ref{Fig_01}.B).

The spatial structure is 
important because of its effect on proximity: 
greater connectedness of a built environment
generally means more direct routes 
and thus shorter distances 
between possible destinations. 
Connectedness also 
affect walking by expanding 
the choice of routes,
 thereby enabling some variety in routes 
within the environment. 
Discovering important spaces of movement
and quantifying differences between them in a spatial graph
of the environment
is not easy since any two spaces can be related 
by means of many paths.
In \cite{Blanchard:2009},
we suggested 
using the properties of  random walks, 
in  order 
to analyze the structure of spatial graphs
and to 
spot structural isolation in
urban environments.
Each node of the graph can be characterized
with respect to the
entire
 graph structure by 
the {\it first-passage time}, 
the expected number of steps 
required to reach the node for the first time 
(without revisiting any intermediate node) starting from 
any node of the graph
 chosen randomly,
 in accordance with the stationary distribution of  random walks. 
In built environments,
the first-passage time to a place 
 can be understood as 
an average number of elementary wayfinding instructions 
(such as "turn left/ right", 
"pass by the door", 
"walk on the corner", etc.)
required to 
  navigate a wanderer to the place
from elsewhere within the environment.
The values of first passage times
to the nodes of the spatial graphs 
A and B are colour indicated (on line)
 in Fig.~\ref{Fig_01}:
 the central places 
characterized with the minimal first passage times 
are red coloured, 
and the secluded places with the maximal first passage times 
are shown in violet. 
The spatial structure of the VE model A
analyzed by means of the spatial graph 
Fig.~\ref{Fig_01}.A
 is essentially 
simpler than the structure of the model B. 
Contrasted to the model A,
where
offices 
are located along three
 sequentially connected 
corridors providing a single path
 between most of possible destinations,
the spatial structure of the model B
allows for many cyclic trips 
due to a network of interconnected 
places and has a number of
vantage points, 
from which a walker
 could observe 
the substantial parts of the environment 
from  different  perspectives (see Sec.~\ref{subsec:reorientations}).
The single central spatial node (the central corridor) 
in the VE model A  is located   
at 9 steps apart
from any randomly chosen node 
in the spatial graph (Fig.~\ref{Fig_01}.A).
The VE model B contains a 
network of well connected spaces 
of movement 
characterized by the first passage times 
ranging from 10 to 14 steps.

\subsection{\label{subsec:Participants} Participants}
\noindent

Two gender balanced groups 
of volunteers
 (82 participants in total)
took part in the controlled  
searching experiments conducted
 in the   office VE
shown in Fig.~\ref{Fig_01}.A - B.
Although participants 
(mostly university students,
with the mean age of  24.2 years 
and a standard deviation of
3.7 years) 
were recruited personally and through advertisements
 at the University of Bielefeld. 
None of them have ever been familiar with the 
actual building prototyping the virtual models A and   B.
Prior to testing, all adult participants
 and parents of children younger than 16 years old
  gave their informed written consent for participation 
in the study. 
Participation in the study was voluntary, and a participant could
revoke her participation consent and quit at any time and for any reason.
The standard provisions for data protection were adhered: 
all test results were kept confidential.
All individual data were managed and processed anonymously
that  eliminated any possibility of identification of participants.

\section{Results and discussions \label{sec:Results}}
\noindent

\subsection{Scanning turns, reorientations, and explorative rotations in VE 
\label{subsec:reorientations}}
\noindent

Self-motion through the VE suffers from a lack of   
many natural body-based cues.
Natural methods of visual
exploration 
 are  also restricted in VE to that
 experienced through the display 
representing only a limited field of view,
suffering from 
the degradation of sensory cues 
due to device latencies
and blocking out all surrounding
visual input.
It was concluded from various experiments
that the 
optic flow without proprioception, 
at least for the limited field of view 
of the virtual display system, 
appears to be
not effective for the updating of heading direction \cite{Klatzky:1998}, 
and 
even when
physical motion cues from free walking are included, 
this is not necessarily sufficient to enable good spatial orientation
in VE \cite{Riecke:2005}.

\begin{figure}[ht!]
 \noindent
\begin{center}
  \epsfig{file=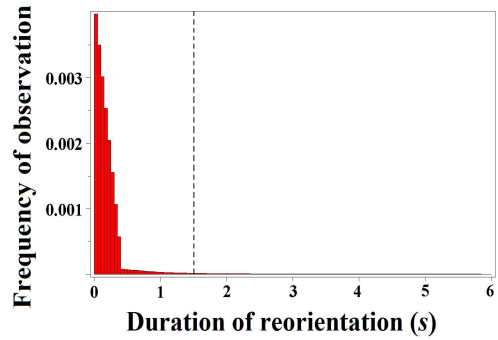, width=6.5cm, height =4.5cm}
\end{center} 
 \caption{ 
The typical distribution of 
scanning turns, reorientations, and explorative rotation
durations while exploring a VE.
The vertical dashed line indicate
 the duration of turn $1.5${\it s}. \label{Fig_02ab}}
\end{figure}

Perhaps due to systematic spatial orientation
 problems occurring in VE  compared to
real-world situations,
the most of subjects  
participated in the our study 
permanently performed fast scanning turns (of $200-300${\it ms})
  by the quick altering pressing
 on the left and right buttons
 of the remote controller, 
each time causing them to turn a greater or lesser angle.
  Probably, such a movement routine
played an important role 
for the proper self-motion perception, as
 compensating  information deficiency experienced
by subjects
while moving through VE under
reduced multisensory conditions.
Longer turns (taking about half a second) 
were observed when subjects redirected their walk or
 avoided obstacles. 
Eventually, the very long, explorative rotations
 often including several complete revolutions
(each time lasting up to a few seconds)
 occurred after far relocations,
 at vantage points,
along the borders of two or several vista spaces,
and at intersections of corridors
that afford a broader view of the environment.  

\begin{figure}[ht!]
 \noindent
\centering
\begin{tabular}{llrr}
1.)& \epsfig{file=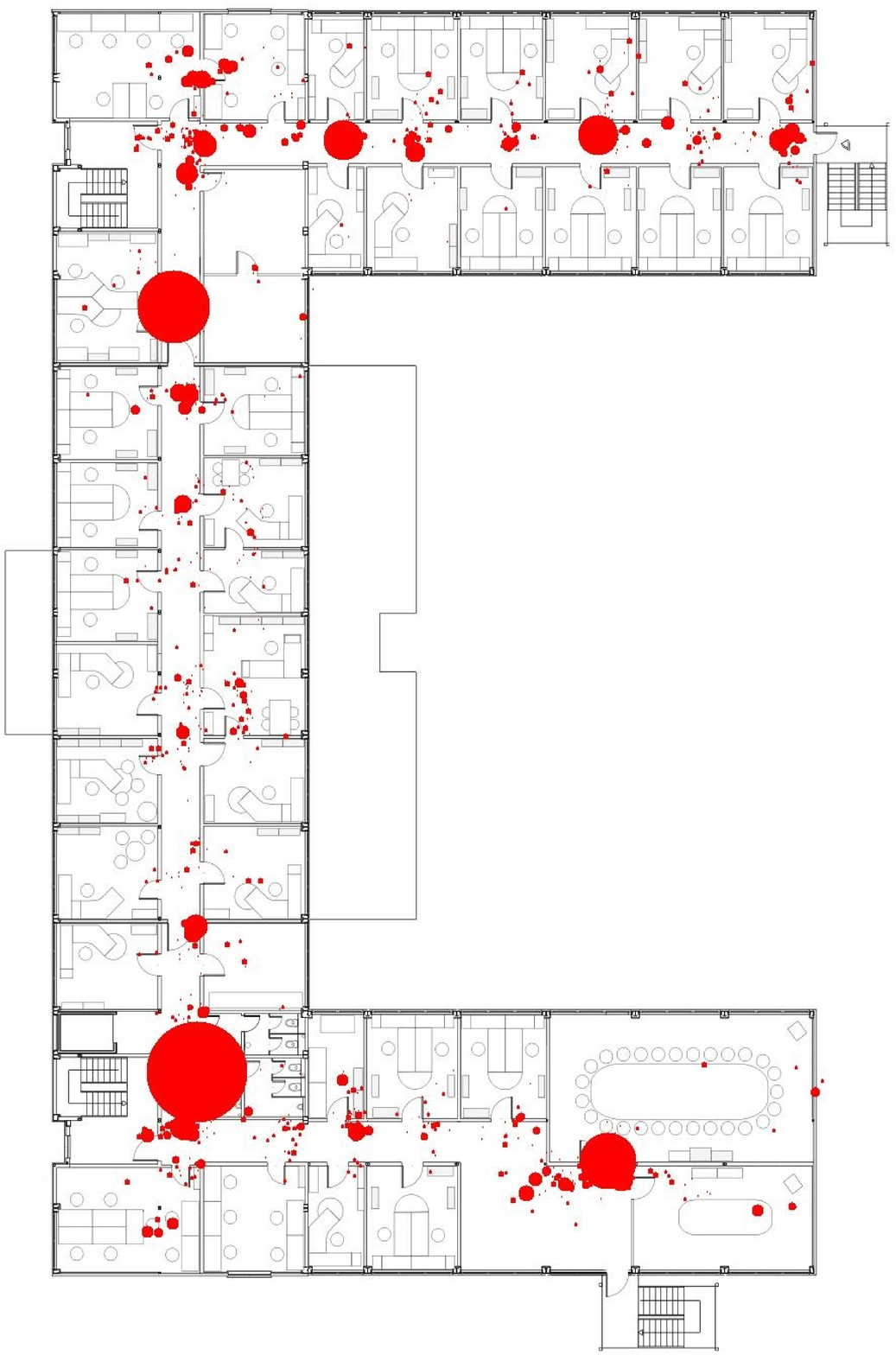, width=4cm, height =6cm} &
2.) &\epsfig{file=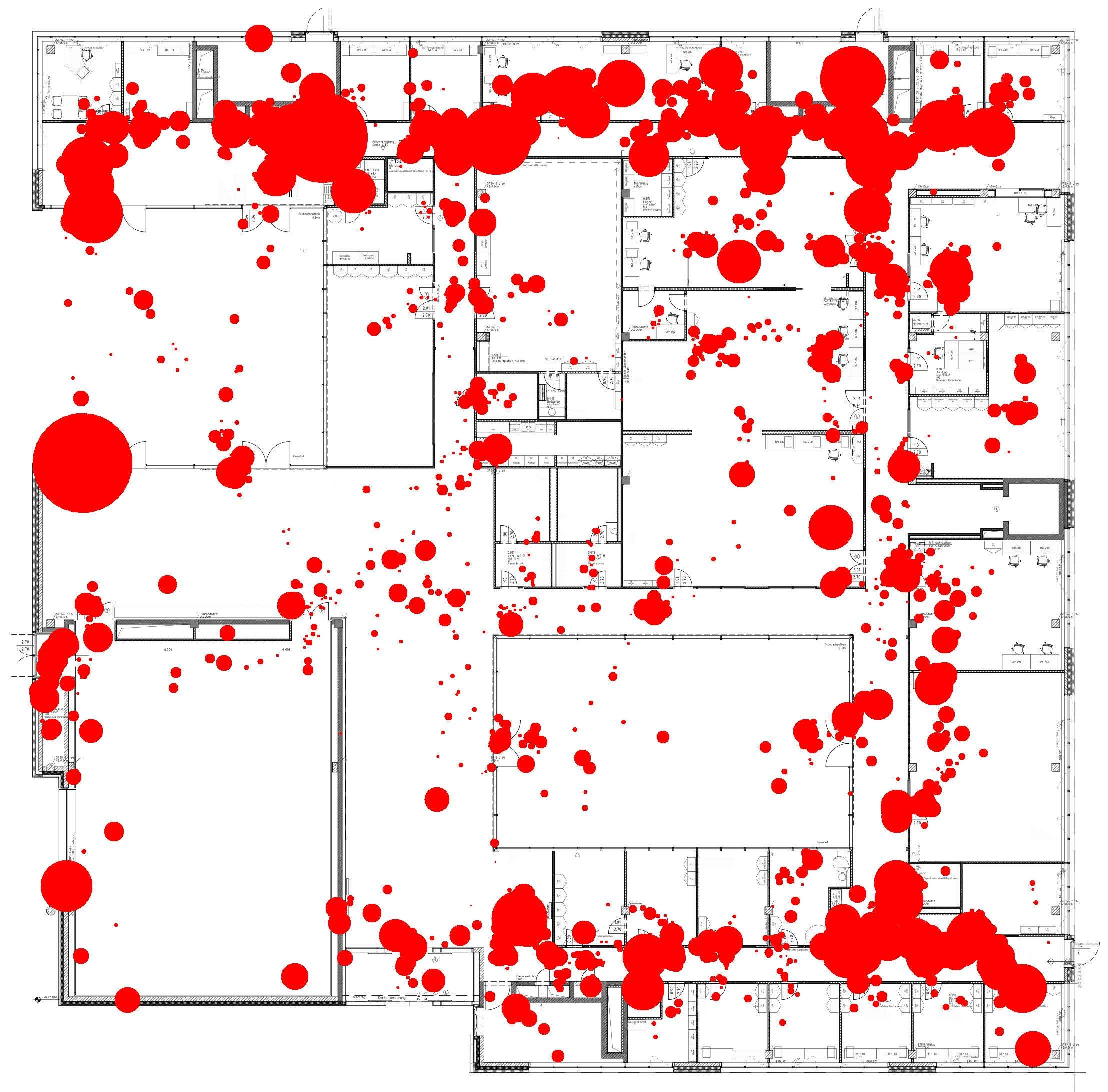, width=6.5cm, height =6cm}
\end{tabular}
\caption{ The floor plans of the  models A and B 
with the locations of vantage points (shown by circles),  
where participants performed reorientations and
 explorative turns lasting  longer than $1.5${\it s}.
  The diameter of a circle 
is proportionate to a number of long turns recorded
at its central point.
 \label{Fig_02a}}
\end{figure}

The typical distribution of
scanning turns, reorientations, and explorative rotation
durations while exploring a VE
is presented on Fig.~\ref{Fig_02ab}.
The areas of 
adjacent rectangles in the histogram are 
equal to the relative frequency of observations in the 
  duration interval. 
 The total area of the histograms 
is normalized to the number of data occurrences. 
The data show that 
the vast majority of all reorientations
performed by subjects 
were the quick scanning turns, 
being the essential part of the adaptive walking routine in VE.
The vertical dashed line in Fig.~\ref{Fig_02ab}
stands for the rotation duration of $1.5${\it s}; 
the locations of the correspondent points, 
at which
participants  performed 
longer turns (without translations) 
are displayed on the floor plans
by the circles
(see Fig.~\ref{Fig_02a}.1 and \ref{Fig_02a}.2).
 The diameter of a circle 
is proportionate to a number of long turns recorded
at its central point over all subjects.

Several studies conducted 
on small animals
\cite{Bartumeus:2008,Bartumeus:2008a},
\cite{Komin:2004,Schimansky-Geier:2005}
 suggested  that 
the switch between scanning and
reorientation behaviour
 in movement patterns of animals  that search 
emerges from complex mechanic-sensorial 
responses of animals to the local environment
and could infer the effects of
 limited perception and/or
a patchy environmental structure.
When exploring patchy resources, 
animals could adjust
turning angle distributions,
selecting a preferred turning
angle value that 
would allow organisms to stay within
the patch for a proper amount of
time, maximizing the energetic gain.
For example, the zigzag motion 
of {\it Daphnia}
appears to be  an
optimal strategy for patch exploitation
  \cite{Schimansky-Geier:2005}.
Therefore,
 the distinction  
between quick scanning turns and a longer reorientation behaviour 
is  crucial to understand the statistical patterns of search
\cite{Bartumeus:2007}.  

\begin{center}
\begin{figure}[ht!]
 \noindent
\begin{center}
\epsfig{file=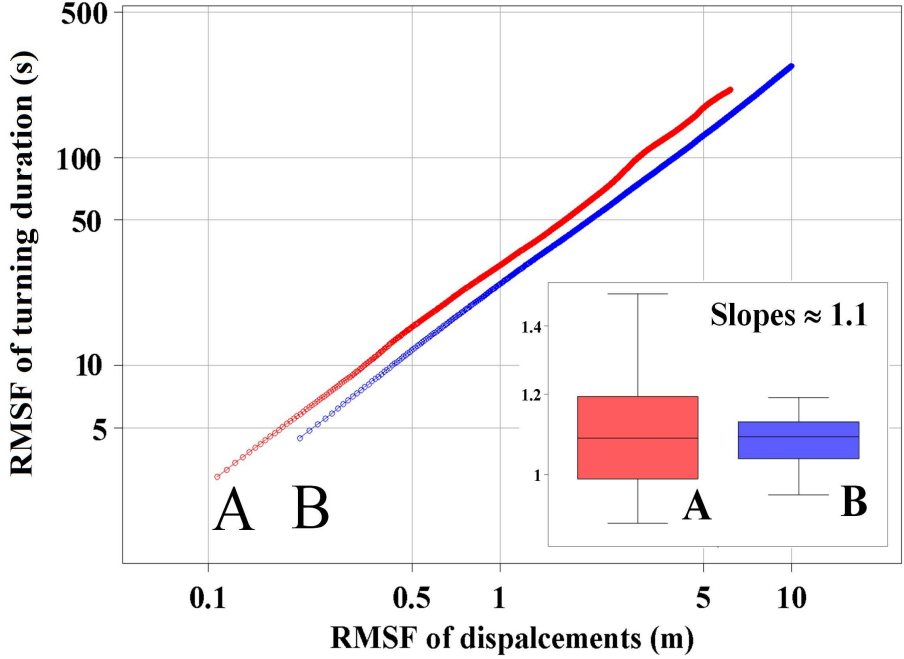, width=12cm, height =9cm} 
\end{center}
\caption{ The root mean square fluctuations (RMSF)
 of the total turning durations   
is shown via the 
RMSF of the net displacement in the log-log scale,
 for all recorded reorientation points of all subjects,
 in the both VE models.
 \label{Fig_05a}}
\end{figure}
\end{center}
Being a part of the travelling routine in  VE, 
quick scanning turns performed by subjects 
during the walk
induce correlations 
between turning 
the total rotation
durations and the net displacements,
on relatively small spatiotemporal scales
sensitive to 
the local structural features
of the environment.
After reaching the natural limits of 
 available space of unobstructed motion 
or entering a new movement zone, 
the subject performs a longer reorientation, 
perhaps in order to explore 
the new environment visually 
that breaks the correlations.  
We analyzed the data 
on durations of reorientations of subjects 
 travelling through the both VE models
 with the use of the {\it root mean square fluctuations} (RMSF)
 suitable for 
detecting correlations \cite{Hurst:1965,Viswanathan:1996}.

The RMSF of displacements is calculated by 
$
F_s(n) = \sqrt{ \left\langle\left(\Delta \mathfrak{S}(n)\right)^2\right\rangle -
\left\langle\Delta \mathfrak{S}(n)\right\rangle^2},  
$
in which 
the net displacement of the walker
by the $n$-th reorientation is
$\mathfrak{S}(n)=\sum_{k=1}^n\left\|{\bf r}_{k+1}-{\bf r}_k\right\|,$ 
${\bf r}_k$ is the recorded position of the $k$-th reorientation,
$\Delta \mathfrak{S}(n)=\mathfrak{S}(n+n_0)-\mathfrak{S}(n_0),$ and 
 the angular brackets denote averaging over all data points
$n_0=1,\ldots n_{\rm max}$.
Similarly, 
the RMSF of rotation durations is calculated by 
$
F_\tau(n) = \sqrt{ \left\langle\left(\Delta \mathfrak{T} (n)\right)^2\right\rangle -
\left\langle\Delta \mathfrak{T} (n)\right\rangle^2},  
$
in which 
the total rotation duration of the subject 
by the $n$-th reorientation is 
$\mathfrak{T}(n)=\sum_{k=1}^n\tau_{k},$ 
$\tau_k$ is the duration of the $k$-th reorientation,
$\Delta \mathfrak{T}(n)=\mathfrak{T}(n+n_0)-\mathfrak{T}(n_0),$ and 
 the angular brackets again denote the average over all data points.
In Fig.~\ref{Fig_05a}, we juxtaposed the RMSF 
of the net displacement and 
the RMSF of the total rotation durations,
 in the log-log scale,
 for all recorded reorientations of subjects,
separately for the VE models A and B.
Identically, for the both VE models,
the graphs show 
the super linear
slope, 
${d\log F_\tau}/{d \log F_s}\cong 1.1,$
indicating the presence of 
a strong positive relation 
that reinforces 
the total duration of quick scanning turns, 
with the increase of the net displacement of subject.
In the long run, the correlations generated by 
the fast scale scanning behaviour vanish 
that is typical for a correlated 
random walk process
\cite{Bartumeus:2008}. 
Interestingly, 
the ranges of 
correlations of 
the RMSF of turning durations
and the RMSF of net displacements 
in the VE models A and B
were different:
for the VE model A, the correlations
exist in the range of the net displacements 
from 0.1 {\it m} to 6 {\it m}, 
and 
 from 0.3 {\it m} to 10 {\it m},
for the VE model B.
Perhaps, the difference
 in the correlation ranges 
of the net displacements 
is due to the different 
 size of an average space
available for unobstructed movements, 
in the models A and B.
It is obvious that such 
an intensive space scan 
is performed by subjects only 
within their immediate neighbourhoods 
and principally cannot be extended 
neither to the entire VE, nor even
to any of its significant parts, 
as the superlinear increase of 
required time 
makes the scanning process
on large spatiotemporal scales
 biologically   
unfeasible. 
Thus, after completing 
a phase of intensive search
 within the patch 
of a size depending on 
the structural properties of the environment,
 a treasure hunter
moves to some other area,
where the phase of intensive search is resumed.
It is important to mention  that 
changes in reorientation behaviour,
on large spatiotemporal scales,
 can generate different anomalous diffusion 
regimes, which in turn, can affect the 
search efficiency of random exploration processes \cite{Bartumeus:2008}.

\subsection{Heavy-tailed distributions of human travels in   VE \label{subsec:Levy}}
\noindent 
 
\begin{figure}[ht!]
\centering
\begin{center}
\epsfig{file=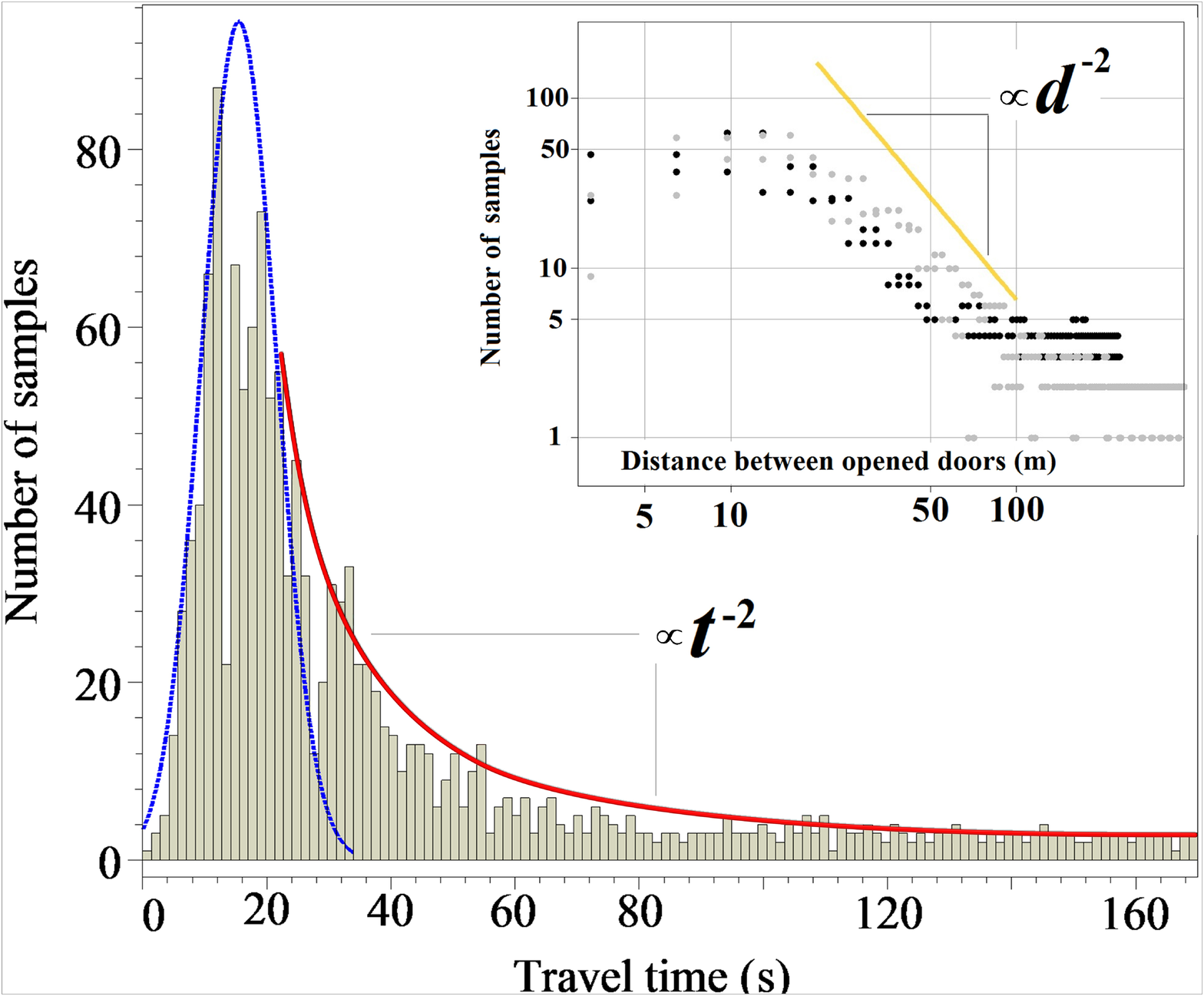, width=12cm, height =9cm} 
\end{center}
\caption{ 
The histogram shows
the frequency of travelling times 
between the consequent door openings as a function of time.
The bell-shaped dotted curve represents the Gaussian 
distribution. 
Occasional long travels contribute
into
the long right tail 
of the distribution
dominated by a quadratic hyperbola.
On the outline, the 
data on dispersal of the consequent searching events 
are shown as a function of distance in the log-log scale
for the VE model A (by black points) and 
for the VE model B (by grey points).
The solid guide line indicates the inverse quadratic slope.
The quadratic  hyperbolic decline constitutes a "fat-tail" effect
detected in the 
foraging movement patterns 
of many living spices,   
 \cite{Viswanathan:1996}-\cite{Buldyrev:2001}, \cite{Shlesinger:1986}-\cite{Sims:2012},
in accordance with the L\'{e}vy foraging hypothesis 
  \cite{Bartumeus:2007}-\cite{Viswanathan:2011}.
 \label{fig1}}
\end{figure}

In order to identify 
phases of specific activity 
in recorded movement patterns and 
to reveal the underlying cognitive mechanisms 
from their statistical properties, 
we have studied the probability distributions
of time intervals and distances between 
consequent observable searching events 
(door openings) as they 
can determine strong changes
in the diffusive properties of movement 
and in relevant spatial
properties of the trajectories.
The distributions representing the data
on dispersal
of the treasure hunters 
 during the experiments 
are shown in Fig.~\ref{fig1}, for both VE models.
Gender is  found as a factor influencing
navigation in VE:
 males were reported to acquire route knowledge from landmarks faster than
females \cite{Cutmore:2000}
 and to spend less time in locating targets \cite{Lin:2012}.
Interestingly, 
the form of the empirically observed 
distributions (Fig.~\ref{fig1}) were neither gender specific,
 nor sensitive to the different global 
structure of the environments.

The data show that 
the most of consequent door openings 
happens in the immediate neighbourhood 
of the actual position of a participant.
The dispersal statistics
 on the small spatiotemporal scales 
shown in Fig.~\ref{fig1} 
can be well approximated 
by uncorrelated Gaussian random walks 
(see the dotted bell shaped curve)
insensitive to the local structure of the environment.
The both distributions
in Fig.~\ref{fig1} 
are remarkable for the long right tails 
dominated by the {\it quadratic hyperbolas} 
attenuating the super-diffusive spread of 
treasure hunters on large spatiotemporal scales.
A power-law tail 
in the probability distributions
of both travelling times and travelled distances 
 could arise from processes, in which 
neither  time, nor
distance
  have a
 specific characteristic scale, so that 
rare but extremely long and far cry,
'explorative'
  travels can occur,
alternating between sequences of many relatively
short, 'exploitative' travels featuring  local searches. 
The power-law tails in 
the dispersal data 
provide an evidence in favour of a 
strong coupling between 
the large-scale 
 movements 
along the paths 
which cannot be viewed at once \cite{Bartumeus:2009}
and the
environmental structure.
As in many empirical phenomena, the tails of the distributions shown in Fig.~\ref{fig1} follow a power law only for values greater than some minimum scales $s_\mathrm{min}$ in space  (about 10{\it m}) and in time (about 20{\it sec}) that is consistent with the conclusion of the previous section about the dynamic formation of immediate neighbourhoods, in which subjects perform intensive scanning. Our observations are in agreement with the hypothesis that both travelling times and travelled distances observed over the group of people participated in the study were drawn from a scale invariant distribution of the form $f(s) \propto (s_\mathrm{min}/s)^2$ and thus do not change  for  $s \gg s_\mathrm{min}$ if scales are multiplied by a common factor. However, it is important to note that statistics of dispersion estimated over a sample of individual searching behaviour could be as different from the power law (\ref{Levy}) as following an exponential distribution, as many of these samples were not statistically representative. Therefore, although the power-law tails of the distributions shown in Fig.~\ref{fig1} are consistent with the L\'{e}vy foraging hypothesis \cite{Bartumeus:2007}-\cite{Viswanathan:2011}, they rather represent a group aggregate feature.

Geometrically, 
discrete-step random walks 
with movement displacements drawn from the 
L\'{e}vy   distribution
  consist of walking clusters with 
very large displacements between them repeated over 
a wide range of scales \cite{Shlesinger:1986,Shlesinger:1993}. 
Although L\'{e}vy flights are ubiquitous for representing 
intermittent search, cruise, and foraging strategies in living spices
to our knowledge, this is the first report 
on the observation of 
L\'{e}vy flight search patterns in humans exploring the VE. 

\section{Model of exploration-exploitation trade-off  for searching behaviour \label{Sec:SOC}}
\noindent

\subsection{Arguments in favor of a self-organized critical model \label{subsec:arguments}}
\noindent

The observed statistical properties 
of human search behaviour in the VE
call for a model 
that could exhibit 
the scale-invariant characteristic 
of a critical point of a phase transition
(correspondent to the super-diffusive scale-free spread
described by the L\'{e}vy distribution)
spontaneously,  
without the need to tune any control parameter to precise values.
In statistical physics, 
 such a property 
of   dynamical systems 
is known as
  self-organized criticality   
\cite{Bak:1987}.
Furthermore, a plausible model has to be of a discrete nature, 
as the biological principle of intermittent locomotion
 assumes that animal behaviour unavoidably produces 
observable punctuations, 
"producing pauses and speeding patterns on the move" \cite{Bartumeus:2008}. 
These motivations reflect a fundamental "trade-off" 
confronted by a treasure hunter choosing
between 
an {\it exploitation} of the scanning movement routine
within the
 familiar environment of nearest neighbourhood,
possibly at no reward, and a fast relocation aiming
 at {\it exploration} of unknown but potentially more 
rewarding areas.

From a theoretical perspective,
it is known that in a stationary setting
there exists an optimal strategy for exploration 
\cite{Gittins:1989}
maximizing the reward 
over an infinite horizon when the value of each reward is discounted exponentially as  
a function of when it is acquired.
However, to date,  
there is no generally optimal solution
 to the exploration versus exploitation 
problem \cite{Cohen:2005,Cohen:2007}, 
as human and other animals are prone to 
{\it dynamically update} their estimates of rewards
in response to diverse, mutable and perhaps discrepant factors,
including elapsed time of search, annoying failures 
to predict the location of a searched object, and  
instantaneous mood swings that could change in a matter of seconds. 
Therefore, it seems that a
stochastic model managing a  balance
between exploration and exploitation 
may be more biologically realistic \cite{Cohen:2007}.
 
There is growing evidence that
the neuromodulatory system
involved in assessing reward and uncertainty
in humans is
 central to the exploration-exploitation trade-off decision \cite{Cohen:2005}.
The  problem can be cast 
in terms of a distinction between 
{\it expected uncertainty}, 
coming from known unreliability of predictive cues and
coded in the brain by a neuromodularity system 
with acetylcholine signals, and 
{\it unexpected uncertainty},
triggered by strongly unexpected observations
promoting exploration and
coded in the brain with norepinephrine signals
 \cite{Yu:2005}.
It was suggested in \cite{Yu:2005} that 
an individual decides on whether 
to stay or to go according  to the current levels
of acetylcholine and norepinephrine,
encoding 
the different types of uncertainty.

Summarizing the above mentioned arguments,
we are interested in a self-organized critical 
 model driven by a discrete time random process 
 of competing between two factors 
featuring the different types of uncertainty.
Various coherent-noise models 
possessing the plausible features
were discussed \cite{Newman:1996,Sneppen:1997}
in connection to the standard sand pile
model \cite{Bak:1987}
 developed in self-organized criticality, 
where the statistics of avalanche
sizes and durations also take a power law form.

\subsection{Mathematical model of decision making in a random search \label{subsec:model}}
\noindent

The movement ecology framework \cite{Nathan:2008}
explicitly recognizes animal movement as a result
of a continuous "dialogue" between environmental cues
(external factors) and animal internal states 
\cite{Bartumeus:2008}. In our model, 
we rationalize the dialogue nature of a 
decision making process 
to take on searching
in highly unpredictable situation 
when no precise information 
on a possibility of rewards is available.
Despite its inherent simplicity, 
the mathematical model 
formulated below  can help generate
hypothesis about fundamental
biological processes and 
 bring a possibility to look for a
variety of biological mechanisms under a common perspective.

We assume that
an individual decides
 on whether 
 to 'exploit'
an immediate neighbourhood 
by
search beyond
 a next door or
to 'explore' 
 other parts
of the environment
yet to be visited
by 
 comparing  the 
 guessed
chances,
$q \in [0, 1]$
 of getting a reward 
beyond the next door 
and $p \in [0, 1]$ of finding 
a treasure elsewhere afar.
It is not necessary that 
$p+q=1.$  
We suppose that 
at each time click 
 subject 
updates one or both 
estimates and 
decides to 
proceed to a part of the environment 
yet to be explored 
if $q < p$.
Otherwise, 
if $q\geq p,$ 
she picks a next door randomly 
among those not yet opened and 
searches  
for a treasure
in the room behind.
We consider 
$p$ and $q$ to be the random variables
 distributed  
over the interval $[0, 1]$ with respect to
the probability distribution functions (pdf)
$\Pr\left\{p<u\right \}=G(u)$ and 
$\Pr\left\{ q<u\right\}=F(u)$ respectively.
In general, $F$ and $G$ are
two arbitrary left-continuous increasing
 functions satisfying the normalization conditions
$F(0) = G(0) = 0$, $F(1) = G(1) = 1$.

We model the intermittent search patterns
by a discrete time random process in the following way. 
At time $t=0,$ the variable
 $q$ is 
chosen with respect to pdf $F$,
 and $p$ can chosen
with respect to pdf $G$. 
If $q < p$, 
subject relocates by pressing a button 
on the controller
and goes to time 
$t = 1$. 
Given a fixed real number 
 $\eta\in [0, 1],$
at time $t\geq  1$,
the following events happen:

(i) with probability $\eta$,
the chance 
 to find a treasure in the immediate neighbourhood
 ($q$) is estimated 
   with pdf $F,$ and the
chance
to get a reward  
elsewhere
($p$) is 
chosen with pdf $G.$

 Otherwise,

(ii) with probability $1-\eta$,
the chance 
to find a treasure in the immediate neighbourhood
($q$) is estimated 
  with pdf $F,$ but the 
chance 
to find it
elsewhere 
($p$)
  keeps
the value it had at time $t- 1$.

Therefore, the parameter $\eta$ quantifies 
the degree of coherence between the two stochastic  sub-processes 
characterized by the pdf $F$ and $G$, respectively: the processes
are coherent if $\eta=1,$ and incoherent 
as $\eta=0.$

If $q\geq p$, the local search phase continues;
however  if $q < p$, 
subject presses the controller button and
moves further, going to time 
 $t + 1$.
Eventually, at some time step $t$, 
when the estimated chance $q$
 exceeds the  value 
$p$,
subject stops and resumes searching 
within the immediate neighbourhood.
The integer value $t = T$ acquired in this random
process limits the  time interval (and travelled distance) 
between sequential phases of searching activity.

\subsection{Analytical solutions for the decision making process \label{subsec:solution}}
\noindent

While studying the  model introduced in Sec.~\ref{subsec:model},
 we are interested in the distribution of   durations of
the relocation phases $P_\eta(T ; F,G)$
 provided the probability distributions $F$ and $G$ are given and the
coherence parameter $\eta$ is fixed.
For many distributions $ F$ and $G$, the model can be solved analytically.
We shall denote $P_\eta(T ; F,G)$ simply by $P(T )$.
A straightforward computation shows directly from the definitions 
 that
$
P(0) =  \int_0^1 dG(p)\left(1-F(p)\right).
$
For $T\geq 1$, the individual can either
depart elsewhere ("{\it D}")
or stay  in the neighbourhood ("{\it S}").
Both events can take place either in the
'correlated' way (with probability $\eta$; see (i)) 
(we denote them $D_c$ and $S_c$), or in
the 'uncorrelated' way (with probability $1 -\eta$; see (ii))
 ($D_u$ and $S_u$). For $T = 1$,
we have for example
\[
\begin{array}{lcl}
P(1) & = & P[DS_c]+P[DS_u] \\
     & = & \int_0^1 dG(p)F(p)\eta(1-F(p)) +
           \int_0^1 dG(p)F(p) (1-\eta)\int_0^1 dG(z)(1-F(z))                              \\
     & = &    \eta B(1) + (1-\eta) A(1)B(0).
\end{array}
\]
Similarly, 
\[
P(2)\,\,=\,\, \eta^2B(2)+\eta(1-\eta)A(1)B(1)+\eta(1-\eta)A(2)B(0)+(1-\eta)^2A(1)^2B(0)
\]
where we define, 
$A(n)=\int_0^1 dG(p)F(p)^n$ and 
$B(n)=  A(n)-A(n+1),$
for $n=0,1,2,\dots.$ 

It is useful to introduce the generating function of $P(T)$,
\[
\hat{P}(s)\,\,=\,\,\sum_{T= 0}^{\infty} s^TP(T),\quad 
P(T)\,\,=\,\,\frac{1}{T!}\left.\frac{d^T \hat{P}(s)}{ds^T}\right|_{s=0}. 
\]
Defining the following auxiliary functions
\[
\begin{array}{lll}
x(l)\,\,=\,\,\eta^l A(l+1), & \mathrm{for}\quad l\geq 1, \quad x(0)=0, \\
y(l)\,\,=\,\, (1-\eta)^l A(1)^{l-1}, & \mathrm{for}\quad l\geq 1, \quad y(0)=0, \\
z(l)\,\,=\,\,\eta^l\left[\eta B(l+1)+ (1-\eta)A(l+1)B(0)\right],
 & \mathrm{for}\quad l\geq 1, \quad z(0)=0, \\
\rho\,\,=\,\,\eta B(1)+(1-\eta)A(1)B(0), & &
\end{array}
\]
we find 
\begin{equation}
\hat{P}(s) = B(0)+\rho s+
\frac{s\left[\hat{z}(s)+\rho \hat{x}(s)\hat{y}(s)+\rho A(1)\hat{y}(s)
+ A(1)\hat{y}(s)\hat{z}(s)\right]}{1-\hat{x}(s)\hat{y}(s)}
\label{eq3}
\end{equation}
where $\hat{x}(s),$ $\hat{y}(s),$ and $\hat{z}(s)$
are   the generating functions of $ {x}(l),$ $ {y}(l),$ and $ {z}(l)$, respectively.
In the marginal cases $\eta = 0$ and $\eta = 1$, the probability $P(T )$ can be readily calculated.
For $\eta = 0$, we have from (\ref{eq3}) 
\begin{equation}
\label{eq4}
\hat{P}_{\eta=0}(s)\,\,=\,\, \frac{B(0)}{1-sA(1)}.
\end{equation}
From  (\ref{eq4}), one gets
\begin{equation}
\label{eq4a}
 {P}_{\eta=0}(T) =  A(1)^TB(0) = \left[
\int_0^1 dG(p)F(p)
\right]^T \int_0^1 dG(p)(1-F(p)).
\end{equation}
Therefore, in this case, for any choice of the pdf $F$ and $G$, the probability $P(T)$ decays
exponentially. 
For $\eta = 1$, (\ref{eq3})  yields
$\hat{P}_{\eta=1}(s)=\hat{B}(s),$ so that 
\begin{equation}
\label{eq5}
P_{\eta=1}(T)\,\,=\,\,B(T)\,\,=\,\,\int_0^1dG(p)F(p)^T(1-F(p)).
\end{equation}
for the special case of uniform
densities $dF(u)  = dG(u) = du$, for all $u \in [0, 1]$ and for
 any $\eta \in [0, 1]$. In this case, simpler
and explicit expressions can be given for $\hat{P}(s)$ and $P(T )$.
Namely, 
from equation (\ref{eq3}), we get 
\begin{equation}
\label{eq6}
\hat{P}(s)\,\,=\,\, \frac 1{1+(1-\eta)\gamma(s)}\left[
\frac{1+\gamma(s)}{s} -\eta \gamma(s)
\right], \quad \gamma(s)\,\,\equiv\,\, \frac{\ln(1-\eta s)}{\eta s}.
\end{equation}
The asymptotic behaviour of $P(T )$ as $T\to \infty$ 
is determined by the singularity of the generating function
$\hat{P}(s)$ that is closest to the origin.
For $\eta = 0$, the generating function 
$\hat{P}(s) = (2 - s)^{-1}$ has a simple pole, and therefore
$P(T )$ decays exponentially that agrees with the result
(\ref{eq4a}).

For the intermediate values $0 < \eta < 1$, the generating function 
$\hat{P}(s)$ has two singularities.
The first pole, $s = s_0$, corresponds to the
 vanishing denominator $1+(1-\eta)\gamma(s)$, where 
$s_0 = s_0(\eta)$
is the unique nontrivial solution of the equation
$ -\ln (1-\eta s)= \eta s/(1-\eta).$
The second singularity, $ s = s_1 =\eta^{-1}$, 
corresponds to the vanishing argument of the logarithm.
It is easy to see that $1 < s_0 < s_1$, so that the 
dominant singularity of $\hat{P}(s)$ is of the polar type, and
for times much larger
than the crossover time $T_c(\eta)\sim \ln \left(s_0(\eta)\right)^{-1}$
the corresponding decay of $P(T $) is exponential,
 with the rate $\ln(s_0(\eta))$.

\begin{figure}[ht]
\noindent
\begin{center}
\epsfig{file=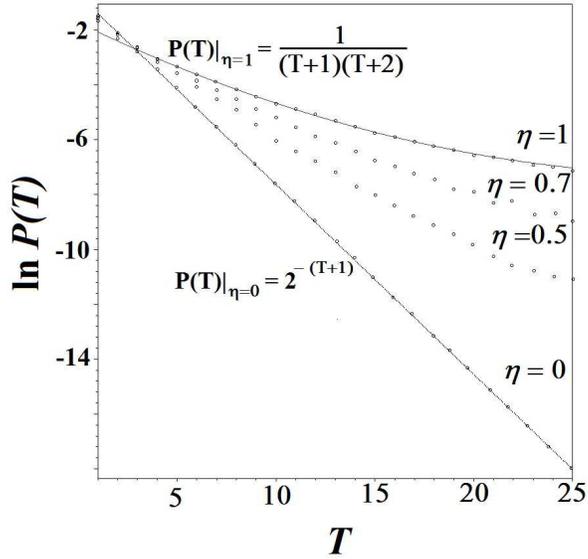, angle= 0,width=8.0cm}
\end{center}
\caption{The distributions of searching durations,
 in the case of the uniform densities $dF(u)\, = \,dG(u)\, =\, du$
  at different values of  $\eta.$ $P_\eta(T )$ 
decays exponentially for $\eta = 0$, 
consistently with the analytical result, $P_{\eta=0}(T ) = 2^{-(T +1)}$ 
(given by the solid line). 
$P_\eta(T )$
exhibits a power-law decay for $\eta\to 1$; 
the solid line is for $P(T)_{\eta=1}=1/(T+1)(T+2)$.\label{fig66}}
\end{figure}

Eventually, 
when $\eta$ tends to $1$, 
the two singularities, $s_0$ and $s_1$ merge. 
More precisely, we have
\begin{equation}
\label{eq11}
\hat{P}_{\eta=1}(s)\,\,=\,\, \frac{s+(1-s)\ln(1-s)}{s^2}.
\end{equation}
The corresponding dominant term in (\ref{eq11}) is of order
$O(T^{-2})$ \cite{Flajolet:2009}. This obviously agrees with
the exact result one can get from equation (\ref{eq5}),
 with $dF(u) = dG(u) = du$,
\begin{equation}
\label{eq12}
P_{\eta=1}(T)\,\,=\,\, \frac 1{(T+1)(T+2)}.
\end{equation}
Let us note that in the case of uniform densities 
it is possible to get an expression of $P_{\eta}(T )$ for all times,
and for any value of $\eta$,
\begin{equation}
\begin{array}{lcl}
P_{\eta}(T) & = & \frac{\eta^T}{(T+1)(T+2)} \\
& & +\sum_{k=1}^T\frac{\eta^T}{k(T-k+1)(T-k+2)}\sum_{m=1}^{k} \left(\frac{1-\eta}{\eta}\right)^m c_{m,k},
\end{array}
\label{eq13}
\end{equation}
in which 
\[
c_{m,k}=m!\sum_{\begin{array}{c}
l_1+\ldots + l_m=k, \\
l_i\ne 0
\end{array} }\frac{l_1l_2\ldots l_{m-1}l_m}{(l_1+1)(k-l_1)\ldots(l_{m-1}+1)(k-l_1-\ldots-l_{m-1})(l_m+1)}.
\]
In Fig.~\ref{fig66}, we have presented 
the probability distributions of the searching durations
for increasing values of $\eta.$
The proposed  mathematical model 
suggests that 
the algebraic tail dominated 
by a 
quadratic hyperbola observed 
 in the 
distribution 
of time intervals and  
between sequential phases of searching activity 
(see Fig.~\ref{fig1})
can arise due to 
 a trade-off between
 exploitation versus exploration 
amid uncertainty.
In such a case, the variable $T$ is set to count bins
 in the histogram shown in Fig.~\ref{fig1} for 
the large scales $s \gg s_\mathrm{min}$.    
 Contrasted to locomotion in real environments,
 mobility in VE depends upon self-motion
perception in virtual space and convenience of 
locomotion interface, rather than on physiological
factors of an individual such as height, weight, age, or fitness. 
The instantaneous translation
velocity in the VE is kept constant for any individual, 
as long as she presses a button.
Therefore the scale-free
distribution 
of time intervals induces
the scale-free
distribution of 
travelling distances, 
with the same scaling exponent.

We have shown 
that 
when 
balancing the chances to be rewarded 
in the immediate neighbourhood ('now and here')
and later elsewhere ('then and there') 
amid  uncertainty
subject
estimates both uniformly at random 
 at each step
($\eta \to 1$),
the inverse quadratic tail 
always dominates the 
bout
distributions 
on large spatiotemporal scales,
so that observed walking behaviour is 
reminiscent of L\'{e}vy flights. 
Thus while searching for a 
sparsely and randomly located objects,
subjects would adopt an 'explorative' movement strategy that takes 
the advantage 
of L\'{e}vy stochastic process, in order to 
minimize the mean time for target detection 
or mean first-passage time to a random target, 
as well as to maximize the energetic gain 
in case of sparsely and randomly 
distributed resources,
since the probability 
of returning to a previously 
visited site is smaller 
 for a L\'{e}vy flight 
than for the usual
random walk 
\cite{Viswanathan:1999}. 
However, if subject is convinced about 
the chance to get rewarded
for exploration keeping this
 estimated value $p$ fixed,
the value of $\eta$ slides from 
$1$ to $0$ (the two stochastic sub-processes 
become incoherent), and the 
walking statistics is tuned out 
from the power law 
(\ref{eq12}) 
to the Brownian walks characterized by 
the exponential decay  
$P_{\eta=0}(T)=2^{-(T+1)}$ (see Fig.~\ref{fig66}).  
For the intermediate values 
of the coherence parameter,
$0 < \eta < 1$,
the saltatory searching behaviour 
possesses both L\'{e}vy and Brownian features.

If we assume that the probability density functions characterizing  the chances for exploration and exploitation in individuals are uniform, $dF(u)=dG(u)=du,$ we can calculate from (\ref{eq13}) the appropriate intermediate value of the coherence parameter $0<\eta<1$ generating the probability distribution $P_{\eta}(T)$ that fits a given empirical distribution $\tilde{P}(T)$ best. The obtained value of $\eta$ can be considered as a measure of instability in subject's estimation of the chance of success. It is also important to mention that the equation $ P_{\eta}(T)= \tilde{P}(T)$ where $ P_{\eta}(T)$ is defined by (\ref{eq13})  can have many solutions within the interval $\eta\in[0,1]$, so that the several consequent values of $T$ have to be checked in order to determine the value of the uncertainty parameter $\eta$ balancing exploration and exploitation behaviours in a searching individual. It is remarkable that the value of $\eta$ can be evaluated formally for a sample of individual searching behaviour even when just a few points are available. However, there are also samples that cannot be fitted reasonably well with (\ref{eq13})  by minimising the discrepancy $\min_{\eta}\sum_{T\geq 1}\left| P_{\eta}(T) -  \tilde{P}(T) \right|$. Perhaps, some individuals exhibit certain preferences while estimating the chances for exploration and exploitation, so that the pdf $F(u)$ and $G(u)$ for them are  not uniform.

\section{Conclusions}
\noindent

We have studied the human
search behaviour in the different office VE. 
The high resolution data on 
displacements and reorientations of 82 subjects 
participated in the treasure hunting experiments
were analysed 
in search of certain statistical regularities. 
The data show that the vast majority of reorientations performed 
by subjects were the quick scanning turns (of 200-- 300 {\it ms}) 
being the essential part of the adaptive movement strategy
under reduced natural multisensory conditions in VE.
The analysis of the root mean square fluctuations
of displacements and turning durations
gives us a conclusive evidence of  
that the total reorientation durations
are  strongly reinforced with the net displacement
 of subjects that makes the intensive scanning process
 biologically unfeasible, 
on large spatiotemporal scales. 
In absence of any specific cues 
marking the location of hidden treasures,
 subjects searched in a saltatory fashion: 
they marched along corridors and across halls,
paused for the local search in the nearest rooms,
and then resumed traversing the VE. 
In built environments, 
the area of an intensive local search is naturally 
limited to 
space available for the unobstructed movement,
in which
 the searching activity 
in humans is characterized 
by the usual diffusive spread.
However, on large scales,
the searching behaviour has the
super-diffusive characteristics: 
we have found that 
the empirical distributions of time intervals 
and distances between consequent searching events 
are dominated by the quadratic hyperbolas
that fits 
the 
L\'{e}vy flight patterns identified
in the intermittent 
search, cruise, and foraging 
behaviours of many living organisms.

Contrary to previous approaches
focused primarily on 
 applying various statistical methods 
for the detection of 
L\'{e}vy flight patterns 
in the available empirical data,
without discussing 
the possible biological mechanisms 
causing the super-diffusive spread 
of searching activity in 
unpredictable environments,
we have suggested a simple stochastic model
of the coherent-noise type
describing 
the exploration-exploitation trade-off 
in humans. 
According  to our model,  
 the saltatory search behaviours
 in humans
 can be featured by 
balancing between exploration and exploitation 
 amid uncertainty
("should I stay or should I go")
 in a way of 
a regularly recurring comparison of 
estimated chances to find a treasure
in an immediate neighbourhood 
and 
to get rewarded 
 in other parts of the environment yet to be explored. 
We have solved the model analytically and
 investigated the statistics of 
possible outcomes of 
such an exploration-exploitation trade-off process.
It is important to mention that 
our model exhibits a variety of saltatory behaviours, 
ranging from L\'{e}vy flights 
occurring under uncertainty 
(when chances 
to get rewarded for the further exploration 
are revaluated by subjects recurrently,  
at each step) 
to Brownian walks, 
with an exponential distribution of movement bouts,
taking place when 
 subject's estimation of the 
chance of success 
remains stable over time
(although not necessary high).
Our model of decision making amid uncertainty
 provides a possible explanation for
the appearance of L\'{e}vy flight 
patterns in the foraging behaviour of animals 
occupying unpredictable environments, such 
as habitats with sparsely distributed resource fields or
in new environments where experience may not help, 
as well as switching between different types of behaviours
toward the more intensive searching strategy that could occur
during a single trip of a treasure hunter confident 
 of the eventual success.

\section*{Acknowledgment}

The treasure hunting experiments have been supported 
by the Cognitive Interaction Technology - Centre of Excellence 
(CITEC, Bielefeld University).
D.V. gratefully acknowledges
the financial support by the project 
{\it MatheMACS}
 ("Mathematics of Multilevel Anticipatory Complex Systems"),
 the grant agreement no. 318723,
funded by the EC Seventh Framework Programme
 FP7-ICT-2011-8.

\section*{Author Contributions}

Conceived and designed the experiments: MT SK DV. 
Performed the experiments: JH. 
Analyzed the data: DV JH.  
Wrote the paper: DV.

\end{document}